\documentstyle[10pt,epsf,epsfig,dp_delphititle]{dp_delphi}
\makeindex
\def\DpPaperGroup{EP}
\def\DpPaperRef{2000-035 }
\def\DpDate{14 February 2000}
\def\DpAuthors{DELPHI Collaboration}
\def\DpSubmit{Accepted by Physics Letters B}
\def\DpTitle{{ W pair production cross-section \\
           and W branching fractions \\
           in {\mbox{\boldmath${\mathrm{e^+ e^-}}$}} interactions at 189 GeV }}
\def\DpComment{  }
\def\DpEMail{ }
\begin{document}
%%%%%%%%%%%%%%%%%%%%%%%%%% They are a problem with Coll.Sty ?
\makeatletter
%\input{dp_system:coll.sty}
% Collapse citation numbers to ranges.  Non-numeric and undefined labels
% are handled.  No sorting is done.  E.g., 1,3,2,3,4,5,foo,1,2,3,?,4,5
% gives 1,3,2-5,foo,1-3,?,4,5
\newcount\@tempcntc
\def\@citex[#1]#2{\if@filesw\immediate\write\@auxout{\string\citation{#2}}\fi
  \@tempcnta\z@\@tempcntb\m@ne\def\@citea{}\@cite{\@for\@citeb:=#2\do
    {\@ifundefined
       {b@\@citeb}{\@citeo\@tempcntb\m@ne\@citea\def\@citea{,}{\bf ?}\@warning
       {Citation `\@citeb' on page \thepage \space undefined}}%
    {\setbox\z@\hbox{\global\@tempcntc0\csname b@\@citeb\endcsname\relax}%
     \ifnum\@tempcntc=\z@ \@citeo\@tempcntb\m@ne
       \@citea\def\@citea{,}\hbox{\csname b@\@citeb\endcsname}%
     \else
      \advance\@tempcntb\@ne
      \ifnum\@tempcntb=\@tempcntc
      \else\advance\@tempcntb\m@ne\@citeo
      \@tempcnta\@tempcntc\@tempcntb\@tempcntc\fi\fi}}\@citeo}{#1}}
\def\@citeo{\ifnum\@tempcnta>\@tempcntb\else\@citea\def\@citea{,}%
  \ifnum\@tempcnta=\@tempcntb\the\@tempcnta\else
   {\advance\@tempcnta\@ne\ifnum\@tempcnta=\@tempcntb \else \def\@citea{--}\fi
    \advance\@tempcnta\m@ne\the\@tempcnta\@citea\the\@tempcntb}\fi\fi}
 
\makeatother
%%%%%%%%%%%%%%%%%%%%%%%%%% ??????????????????????????????????
% Generate the title page
\begin{titlepage}
\pagenumbering{roman}
\CERNpreprint{\DpPaperGroup}{\DpPaperRef} % Reference of the paper
\date{{\small\DpDate}} % Date of the paper
\title{\DpTitle} % Title of the paper
\address{\DpAuthors} % General name of the author(s)
\begin{shortabs} % Start the abstract
\noindent
\noindent

The cross-section for the process $e^+e^-\rightarrow W^+W^-$ 
has been measured with the data sample collected by DELPHI 
at an average centre-of-mass energy of 189~GeV and
corresponding to an integrated luminosity of 155~pb$^{-1}$. 
Based on the 2392 events selected as $WW$ candidates, the cross-section for the
doubly resonant process $e^+e^-\rightarrow W^+W^-$ has been measured to be
$15.83 \pm 0.38~\mbox{(stat)} \pm 0.20~\mbox{(syst)~pb}$.
The branching fractions of the $W$ decay were also measured and found 
to be in good agreement with the Standard Model expectation.
From these a value of the CKM mixing matrix element 
$ |V_{cs}| = 1.001 \pm 0.040~\mbox{(stat)} \pm 0.020~\mbox{(syst)}$
was derived. 
\end{shortabs}
\vfill
\begin{center}
\DpSubmit \ \\ % Horrible hack to allow to have DpSubmit empty
\DpComment \ \\
\DpEMail \ \\
\end{center}
\vfill
\clearpage
\headsep 10.0pt
\addtolength{\textheight}{10mm}
\addtolength{\footskip}{-5mm}
\begingroup
% Commands to process the author names
%
\newcommand{\DpName}[2]{\hbox{#1$^{\ref{#2}}$},\hfill}
\newcommand{\DpNameTwo}[3]{\hbox{#1$^{\ref{#2},\ref{#3}}$},\hfill}
\newcommand{\DpNameThree}[4]{\hbox{#1$^{\ref{#2},\ref{#3},\ref{#4}}$},\hfill}
\newskip\Bigfill \Bigfill = 0pt plus 1000fill
\newcommand{\DpNameLast}[2]{\hbox{#1$^{\ref{#2}}$}\hspace{\Bigfill}}
%
%\small
\footnotesize
\noindent
\DpName{P.Abreu}{LIP}
\DpName{W.Adam}{VIENNA}
\DpName{T.Adye}{RAL}
\DpName{P.Adzic}{DEMOKRITOS}
\DpName{Z.Albrecht}{KARLSRUHE}
\DpName{T.Alderweireld}{AIM}
\DpName{G.D.Alekseev}{JINR}
\DpName{R.Alemany}{VALENCIA}
\DpName{T.Allmendinger}{KARLSRUHE}
\DpName{P.P.Allport}{LIVERPOOL}
% ...
\normalsize
\endgroup
\titlefoot{Department of Physics and Astronomy, Iowa State
     University, Ames IA 50011-3160, USA
    \label{AMES}}
\titlefoot{Physics Department, Univ. Instelling Antwerpen,
     Universiteitsplein 1, B-2610 Antwerpen, Belgium \\
     \indent~~and IIHE, ULB-VUB,
     Pleinlaan 2, B-1050 Brussels, Belgium \\
     \indent~~and Facult\'e des Sciences,
     Univ. de l'Etat Mons, Av. Maistriau 19, B-7000 Mons, Belgium
    \label{AIM}}
\titlefoot{Physics Laboratory, University of Athens, Solonos Str.
     104, GR-10680 Athens, Greece
    \label{ATHENS}}
\titlefoot{Department of Physics, University of Bergen,
     All\'egaten 55, NO-5007 Bergen, Norway
    \label{BERGEN}}
\titlefoot{Dipartimento di Fisica, Universit\`a di Bologna and INFN,
     Via Irnerio 46, IT-40126 Bologna, Italy
    \label{BOLOGNA}}
\titlefoot{Centro Brasileiro de Pesquisas F\'{\i}sicas, rua Xavier Sigaud 150,
     BR-22290 Rio de Janeiro, Brazil \\
     \indent~~and Depto. de F\'{\i}sica, Pont. Univ. Cat\'olica,
     C.P. 38071 BR-22453 Rio de Janeiro, Brazil \\
     \indent~~and Inst. de F\'{\i}sica, Univ. Estadual do Rio de Janeiro,
     rua S\~{a}o Francisco Xavier 524, Rio de Janeiro, Brazil
    \label{BRASIL}}
\titlefoot{Comenius University, Faculty of Mathematics and Physics,
     Mlynska Dolina, SK-84215 Bratislava, Slovakia
    \label{BRATISLAVA}}
\titlefoot{Coll\`ege de France, Lab. de Physique Corpusculaire, IN2P3-CNRS,
     FR-75231 Paris Cedex 05, France
    \label{CDF}}
\titlefoot{CERN, CH-1211 Geneva 23, Switzerland
    \label{CERN}}
\titlefoot{Institut de Recherches Subatomiques, IN2P3 - CNRS/ULP - BP20,
     FR-67037 Strasbourg Cedex, France
    \label{CRN}}
\titlefoot{Now at DESY-Zeuthen, Platanenallee 6, D-15735 Zeuthen, Germany
    \label{DESY}}
\titlefoot{Institute of Nuclear Physics, N.C.S.R. Demokritos,
     P.O. Box 60228, GR-15310 Athens, Greece
    \label{DEMOKRITOS}}
\titlefoot{FZU, Inst. of Phys. of the C.A.S. High Energy Physics Division,
     Na Slovance 2, CZ-180 40, Praha 8, Czech Republic
    \label{FZU}}
\titlefoot{Dipartimento di Fisica, Universit\`a di Genova and INFN,
     Via Dodecaneso 33, IT-16146 Genova, Italy
    \label{GENOVA}}
\titlefoot{Institut des Sciences Nucl\'eaires, IN2P3-CNRS, Universit\'e
     de Grenoble 1, FR-38026 Grenoble Cedex, France
    \label{GRENOBLE}}
\titlefoot{Helsinki Institute of Physics, HIP,
     P.O. Box 9, FI-00014 Helsinki, Finland
    \label{HELSINKI}}
\titlefoot{Joint Institute for Nuclear Research, Dubna, Head Post
     Office, P.O. Box 79, RU-101 000 Moscow, Russian Federation
    \label{JINR}}
\titlefoot{Institut f\"ur Experimentelle Kernphysik,
     Universit\"at Karlsruhe, Postfach 6980, DE-76128 Karlsruhe,
     Germany
    \label{KARLSRUHE}}
\titlefoot{Institute of Nuclear Physics and University of Mining and Metalurgy,
     Ul. Kawiory 26a, PL-30055 Krakow, Poland
    \label{KRAKOW}}
\titlefoot{Universit\'e de Paris-Sud, Lab. de l'Acc\'el\'erateur
     Lin\'eaire, IN2P3-CNRS, B\^{a}t. 200, FR-91405 Orsay Cedex, France
    \label{LAL}}
\titlefoot{School of Physics and Chemistry, University of Lancaster,
     Lancaster LA1 4YB, UK
    \label{LANCASTER}}
\titlefoot{LIP, IST, FCUL - Av. Elias Garcia, 14-$1^{o}$,
     PT-1000 Lisboa Codex, Portugal
    \label{LIP}}
\titlefoot{Department of Physics, University of Liverpool, P.O.
     Box 147, Liverpool L69 3BX, UK
    \label{LIVERPOOL}}
\titlefoot{LPNHE, IN2P3-CNRS, Univ.~Paris VI et VII, Tour 33 (RdC),
     4 place Jussieu, FR-75252 Paris Cedex 05, France
    \label{LPNHE}}
\titlefoot{Department of Physics, University of Lund,
     S\"olvegatan 14, SE-223 63 Lund, Sweden
    \label{LUND}}
\titlefoot{Universit\'e Claude Bernard de Lyon, IPNL, IN2P3-CNRS,
     FR-69622 Villeurbanne Cedex, France
    \label{LYON}}
\titlefoot{Univ. d'Aix - Marseille II - CPP, IN2P3-CNRS,
     FR-13288 Marseille Cedex 09, France
    \label{MARSEILLE}}
\titlefoot{Dipartimento di Fisica, Universit\`a di Milano and INFN-MILANO,
     Via Celoria 16, IT-20133 Milan, Italy
    \label{MILANO}}
\titlefoot{Dipartimento di Fisica, Univ. di Milano-Bicocca and
     INFN-MILANO, Piazza delle Scienze 2, IT-20126 Milan, Italy
    \label{MILANO2}}
\titlefoot{Niels Bohr Institute, Blegdamsvej 17,
     DK-2100 Copenhagen {\O}, Denmark
    \label{NBI}}
\titlefoot{IPNP of MFF, Charles Univ., Areal MFF,
     V Holesovickach 2, CZ-180 00, Praha 8, Czech Republic
    \label{NC}}
\titlefoot{NIKHEF, Postbus 41882, NL-1009 DB
     Amsterdam, The Netherlands
    \label{NIKHEF}}
\titlefoot{National Technical University, Physics Department,
     Zografou Campus, GR-15773 Athens, Greece
    \label{NTU-ATHENS}}
\titlefoot{Physics Department, University of Oslo, Blindern,
     NO-1000 Oslo 3, Norway
    \label{OSLO}}
\titlefoot{Dpto. Fisica, Univ. Oviedo, Avda. Calvo Sotelo
     s/n, ES-33007 Oviedo, Spain
    \label{OVIEDO}}
\titlefoot{Department of Physics, University of Oxford,
     Keble Road, Oxford OX1 3RH, UK
    \label{OXFORD}}
\titlefoot{Dipartimento di Fisica, Universit\`a di Padova and
     INFN, Via Marzolo 8, IT-35131 Padua, Italy
    \label{PADOVA}}
\titlefoot{Rutherford Appleton Laboratory, Chilton, Didcot
     OX11 OQX, UK
    \label{RAL}}
\titlefoot{Dipartimento di Fisica, Universit\`a di Roma II and
     INFN, Tor Vergata, IT-00173 Rome, Italy
    \label{ROMA2}}
\titlefoot{Dipartimento di Fisica, Universit\`a di Roma III and
     INFN, Via della Vasca Navale 84, IT-00146 Rome, Italy
    \label{ROMA3}}
\titlefoot{DAPNIA/Service de Physique des Particules,
     CEA-Saclay, FR-91191 Gif-sur-Yvette Cedex, France
    \label{SACLAY}}
\titlefoot{Instituto de Fisica de Cantabria (CSIC-UC), Avda.
     los Castros s/n, ES-39006 Santander, Spain
    \label{SANTANDER}}
\titlefoot{Dipartimento di Fisica, Universit\`a degli Studi di Roma
     La Sapienza, Piazzale Aldo Moro 2, IT-00185 Rome, Italy
    \label{SAPIENZA}}
\titlefoot{Inst. for High Energy Physics, Serpukov
     P.O. Box 35, Protvino, (Moscow Region), Russian Federation
    \label{SERPUKHOV}}
\titlefoot{J. Stefan Institute, Jamova 39, SI-1000 Ljubljana, Slovenia
     and Laboratory for Astroparticle Physics,\\
     \indent~~Nova Gorica Polytechnic, Kostanjeviska 16a, SI-5000 Nova Gorica, Slovenia, \\
     \indent~~and Department of Physics, University of Ljubljana,
     SI-1000 Ljubljana, Slovenia
    \label{SLOVENIJA}}
\titlefoot{Fysikum, Stockholm University,
     Box 6730, SE-113 85 Stockholm, Sweden
    \label{STOCKHOLM}}
\titlefoot{Dipartimento di Fisica Sperimentale, Universit\`a di
     Torino and INFN, Via P. Giuria 1, IT-10125 Turin, Italy
    \label{TORINO}}
\titlefoot{Dipartimento di Fisica, Universit\`a di Trieste and
     INFN, Via A. Valerio 2, IT-34127 Trieste, Italy \\
     \indent~~and Istituto di Fisica, Universit\`a di Udine,
     IT-33100 Udine, Italy
    \label{TU}}
\titlefoot{Univ. Federal do Rio de Janeiro, C.P. 68528
     Cidade Univ., Ilha do Fund\~ao
     BR-21945-970 Rio de Janeiro, Brazil
    \label{UFRJ}}
\titlefoot{Department of Radiation Sciences, University of
     Uppsala, P.O. Box 535, SE-751 21 Uppsala, Sweden
    \label{UPPSALA}}
\titlefoot{IFIC, Valencia-CSIC, and D.F.A.M.N., U. de Valencia,
     Avda. Dr. Moliner 50, ES-46100 Burjassot (Valencia), Spain
    \label{VALENCIA}}
\titlefoot{Institut f\"ur Hochenergiephysik, \"Osterr. Akad.
     d. Wissensch., Nikolsdorfergasse 18, AT-1050 Vienna, Austria
    \label{VIENNA}}
\titlefoot{Inst. Nuclear Studies and University of Warsaw, Ul.
     Hoza 69, PL-00681 Warsaw, Poland
    \label{WARSZAWA}}
\titlefoot{Fachbereich Physik, University of Wuppertal, Postfach
     100 127, DE-42097 Wuppertal, Germany
    \label{WUPPERTAL}}
\addtolength{\textheight}{-10mm}
\addtolength{\footskip}{5mm}
\clearpage
\headsep 30.0pt
\end{titlepage}
%%%%%%%%%%%%%%%%%%%%%%%%%
%
% Change for the document body
%%\pagestyle{heading} % for page numbering
\pagenumbering{arabic} % page numbering in number
\setcounter{footnote}{0} %
\large
%\linenumbers %%%CD
\newcommand{\ee}      {\mbox{${{e^+ e^-}}                 $}}
\newcommand{\qq}      {\mbox{${q\bar{q}}                $}}
\newcommand{\qqga}    {\mbox{${q\bar{q}(\gamma)}                $}}
\newcommand{\qqgg}    {\mbox{${q\bar{q}gg}                $}}
\newcommand{\qqenu}   {\mbox{$ \qq e \nu                                   $}}
\newcommand{\qqmunu}  {\mbox{$ \qq \mu \nu                                 $}}
\newcommand{\qqtaunu} {\mbox{$ \qq \tau \nu                                $}}
\newcommand{\jjenu}   {\mbox{$ j j e \nu                                   $}}
\newcommand{\jjmunu}  {\mbox{$ j j \mu \nu                                 $}}
\newcommand{\jjtaunu} {\mbox{$ j j \tau \nu                                $}}
\newcommand{\jjjj}    {\mbox{$ j j j j                                     $}}
\newcommand{\ra}{\rightarrow}
\renewcommand{\deg}{^\circ}
\renewcommand{\textfraction}{0.01}

\section{Introduction}
The cross-section for the doubly resonant production of $W$ bosons has been
measured with the data sample collected by DELPHI at 
the average centre-of-mass energy of $188.63 \pm 0.04$~GeV.
Depending on the decay mode of each $W$ boson, fully hadronic, mixed
hadronic-leptonic (``semileptonic'') or fully leptonic 
final states were obtained, for which the Standard Model branching 
fractions are 45.6\%, 43.9\% and 10.5\%, respectively. 
The detector was essentially unchanged compared to
previous years and detailed descriptions of the DELPHI apparatus and its
performance can be found in \cite{DET,PERF}.
The luminosity was measured using the Small Angle Tile Calorimeter \cite{STIC}.
The total integrated luminosity corresponds to 155~pb$^{-1}$;
its systematic error is estimated to be $\pm 0.6\%$, which is dominated 
by the experimental uncertainty on the Bhabha measurements of $\pm 0.5\%$.
The luminosities used for the different selections correspond to those data
for which all elements of the detector essential to each specific analysis
were fully functional. 
The criteria for the selection of $WW$ events are reviewed in section 2. 
Generally they follow those used for the cross-section measurements at lower 
centre-of-mass energies \cite{wwpap161,wwpap172,wwpap183}, but for the 
4-jet final state a more efficient selection has been applied using a neural 
network, and the selection of leptons has been modified to improve the 
efficiency for $\tau$ leptons.
In section 3 the total cross-section and the branching 
fractions of the $W$ boson are presented.

The cross-sections determined in this analysis correspond 
to $W$~pair production through the three doubly resonant tree-level diagrams 
(``CC03 diagrams''~\cite{CC03}) involving $s$-channel $\gamma$ and $Z$ exchange
and $t$-channel $\nu$ exchange. The selection efficiencies were defined 
with respect to these diagrams only and were determined using the full
simulation program DELSIM \cite{DELSIM} with the PYTHIA 5.7 
event generator~\cite{PYTHIA}.
In addition to the production via the CC03 diagrams, the
four-fermion final states corresponding to some decay modes may be 
produced via other Standard Model diagrams involving either zero, one, or two 
massive vector bosons. Corrections which account for the interference between 
the CC03 diagrams and the additional diagrams are generally expected to be
negligible at this energy, except for final states with electrons or positrons.
In these cases correction factors were determined from simulation
using the 4-fermion generator EXCALIBUR~\cite{EXCALIBUR} and were found 
to be consistent with unity within an uncertainty of $\pm 2$\%.

\section{Event selection and cross-sections}
\label{sec:cs}

\subsection{Fully hadronic final state}

A feed-forward neural network (see e.g. \cite{rojas,bishop}) was
used to improve the selection quality of $W^{+} W^{-} \rightarrow q
\overline{q} q \overline{q}$ from 2-fermion (mainly $Z/\gamma
\rightarrow q \overline{q}$) and 4-fermion background (mainly $Z Z
\rightarrow q\bar{q}x\bar{x}$). Compared to an analysis based on sequential
cuts \cite{wwpap183}, the selection efficiency for the signal was 
increased by about 10\% for the same purity. 
The network is based on the JETNET package
\cite{jetnet}, uses the standard back-propagation algorithm, and consists
of three layers with 13 input nodes, 7 hidden nodes and one output node. 

A preselection of the events was performed with the following criteria:
\begin{itemize}
\item{a reconstructed effective centre-of-mass energy \cite{SPRIME} 
$\sqrt{s'} > 115$~GeV;}
\item{4 or more reconstructed jets when clustering with LUCLUS \cite{LUCLUS} 
at $d_{join} = 4.0~{\mathrm GeV}/c$;}
\item{total particle multiplicity $\geq 3$ for each jet.}
\end{itemize}
Each event was forced into a 4-jet configuration.
The following jet and event observables were chosen as input variables,
taking into account previous neural network studies to optimize input variables
for the $WW$ and 2-fermion separation \cite{somap}:
\begin{enumerate}
\item{the difference between the maximum and minimum jet energy after
      a 4C fit, imposing 4-momentum conservation on the event;}
\item{the minimum angle between two jets after the 4C fit;}
\item{the value of $d_{join}$ from the cluster algorithm LUCLUS
for the migration of 4 jets into 3 jets;}
\item{the minimum particle multiplicity of all jets;}
\item{the reconstructed effective centre-of-mass energy $\sqrt{s'}$;}
\item{the maximum probability (for all possible jet pairings) for a 2C fit
(two objects with $W$-mass);}
\item{thrust;}
\item{sphericity;}
\item{the mean rapidity of all particles with respect to the thrust axis;}
\item{the sum of the cubes of the magnitudes of the momenta of the 7 highest 
      momentum particles $\sum_{i=1}^{7} |\vec{p}_{i}|^{3}$;}
\item{the minimum jet broadening $B_{min}$ \cite{catani};}
\item{the Fox-Wolfram-moment H3 \cite{fox};}
\item{the Fox-Wolfram-moment H4.}
\end{enumerate}

\begin{figure}[htb]
\centerline{\epsfig{file=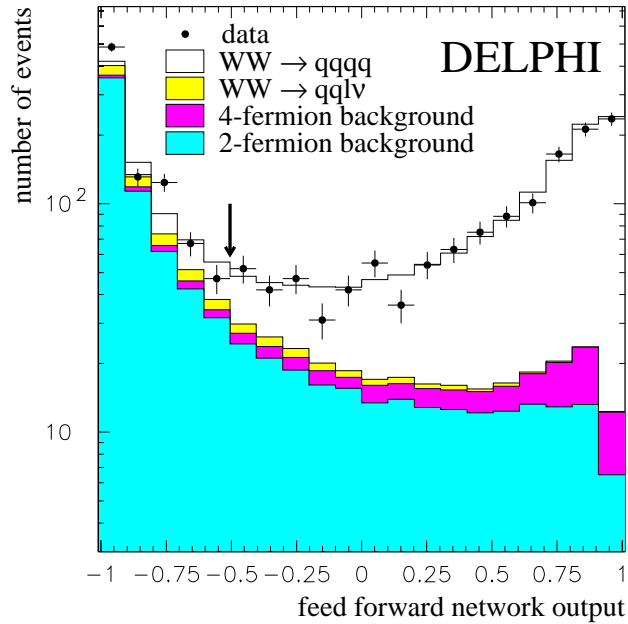,width=10cm}}
\caption{
Distribution of the neural network output variable for 4-jet
events. The points show the data and the histograms are
the predicted distributions for signal and background. Events to the right of
the vertical arrow were accepted for the event sample.}
\label{fig1} \end{figure}

The training of the neural network was performed with 3500 signal events
($WW\ra\qq\qq$) and 3500 $Z/\gamma\ra\qq $ background events simulated 
with the PYTHIA 5.7 event generator. Afterwards the network output was
calculated for other independent samples of simulated $WW$, $Z/\gamma$ 
and $ZZ$ events and for the real data. Figure \ref{fig1} shows the 
distribution of the neural network output for data and simulated events.

For each bin the fractional efficiencies of the selection of $WW$
decays and the background contributions were estimated from simulation. 
Events were finally retained if the neural network output variable was larger 
than -0.5.
The resulting selection efficiencies for the $WW$ channels are listed 
in the second column of Table~\ref{qqxx} together with the estimated 
backgrounds.

\begin{table}[htb]
\begin{center}
\begin{tabular}{|c|c|c|c|c|c|c|}
\hline
& \multicolumn{4}{|c|}{efficiencies for selected channels} \\
\hline
channel              & $\jjjj$ & $\jjenu$ & $\jjmunu$ & $\jjtaunu$ \\
\hline
$\qq\qq$      & {\bf0.887} & 0.         & 0.         & 0.018 \\
$\qqenu$      & 0.009      & {\bf0.661} & 0.         & 0.115 \\
$\qqmunu$     & 0.004      & 0.         & {\bf0.865} & 0.056 \\
$\qqtaunu$    & 0.028      & 0.039      & 0.034      & {\bf0.491} \\
\hline
background (pb)  & 1.788 & 0.159 & 0.075 & 0.437 \\
selected events  & 1298  & 259   & 328   & 324   \\
\hline
luminosity (pb$^{-1}$)      & \multicolumn{1}{|c|}{154.35} 
                            & \multicolumn{3}{|c|}{153.94} \\
\hline
\end{tabular}
\vspace{0.2cm}
\caption{\label{qqxx} Selection efficiencies, background and data for 
the hadronic and semileptonic final states.}
\end{center}
\end{table}

A relative uncertainty on the efficiency of $\pm 0.6\%$ was estimated 
from the following studies:
\begin{itemize}
\item
{Comparison of the simulation with real data, which were taken at a 
centre-of-mass energy of 91.2~GeV with the same detector and trigger 
configuration and analysed with the same reconstruction software 
as the 189~GeV data.   
For this comparison, the technique of mixing Lorentz-boosted 
$Z$ events was used, which transformed two independent hadronic $Z$ decays into
a pseudo $W$ pair event by applying an appropriate boost to the particles
of each $Z$ decay. In this study a difference of 0.17\% was observed.}
\item
{Variation of the efficiency using different hadronization models 
(JETSET 7.4 \cite{PYTHIA} and ARIADNE \cite{ARIADNE}) giving 0.59\%.
Within this error variations of the efficiency from different modelling 
of Bose-Einstein correlations in the generator were found to be negligable; 
this is also expected for final state interactions between quarks from 
different W bosons (``Colour Reconnection'').}
\end{itemize}

The cross-section for the expected total background was estimated from
the simulations to be $(1.79 \pm 0.09)$~pb. The main contribution
(1.42~pb) comes from $\qqga$ events with gluon radiation, 
the rest from non-$WW$ 4-fermion final states.
The systematic uncertainty on the background was estimated from the 
variation of the selection efficiencies for the different backgrounds when 
different hadronization models were used (JETSET 7.4 and ARIADNE). 
A total uncertainty of $\pm 5\%$ was estimated from this variation (4.5\%) and
from differences between data and simulation due to imperfections of the
generator models.
Furthermore, the influences of the different parameters of the neural network 
structure, its learning algorithm, and of the preselection have been 
investigated and found to be negligible.

A total of 1298 events were selected in the data sample. The cross-section 
for fully hadronic events was obtained from a binned maximum
likelihood fit to the distribution of the neural network output variable
above -0.5, taking into account the expected background in each bin. 
The result is

$$ \sigma_{WW}^{qqqq}= \sigma_{WW}^{tot} \times 
   {\mathrm{BR}}(WW\rightarrow \qq\qq) 
   = 7.36  \pm 0.26~\mbox{(stat)}  \pm 0.10~\mbox{(syst)} ~~\mbox{pb}, $$

\noindent where BR$(WW\rightarrow \qq\qq)$
is the probability for 
the $WW$ pair to give a purely hadronic final state. The first error 
is statistical and the second is systematic.  The systematic error 
includes contributions from the uncertainties on efficiency and background, 
and on the luminosity.

\subsection{Semileptonic final state}   

Events in which one of the $W$ bosons decays into $l\nu$ and the other one into 
quarks are characterized by two hadronic jets, one isolated lepton 
(coming either directly from the $W$ decay or from the cascade decay 
$W \rightarrow \tau\nu \rightarrow e\nu\nu\nu \: {\mathrm or} \: \mu\nu\nu\nu$)
or a low multiplicity jet due to a
$\tau$ decay, and missing momentum resulting from the neutrino.  
The major background comes from $\qqga$ production and from four-fermion 
final states containing two quarks and two leptons of the same flavour.

Events were required to show hadronic activity (at least 6 charged particles),
to have a total visible energy of at least 50~GeV and to be compatible
with a 3-jet topology on application of the LUCLUS~\cite{LUCLUS} clustering
algorithm with a value of $d_{join}$ between 2 and 20~GeV/$c$.
A lepton had to be found in the event according to one of the following 
criteria:
\begin{itemize}
\item {a single charged particle identified as described in \cite{wwpap172}
as an electron or a muon, and with at least 5~GeV of energy;}
%(the identification is performed in the same way 
%as described in \cite{wwpap172});}
\item {a single charged particle of momentum, $p$, above 5~GeV/$c$, 
not identified as a lepton, but isolated from other 
particles by $p\cdot\theta_{iso}>100~{\mathrm GeV}/c\cdot$degrees,
where $\theta_{iso}$ is the angle formed with the closest charged particle
with a momentum of at least $1~{\mathrm GeV}/c$;}
\item {a low multiplicity jet (with less than 4 charged particles) with 
an energy above 5~GeV, polar angle of its axis with respect to the beam axis 
between $20\deg$ and $160\deg$, 
and with at least 15\% of its energy carried by charged particles.}
% The
%fraction of the charged energy of this candidate $\tau$ jet with respect to 
%its total energy was required to exceed 15\%.}
\end{itemize} 
When several candidate leptons with the same flavour were found in the 
event, the one with highest $p\cdot\theta_{iso}$ (single-track case) 
or smallest opening angle (jet case) was chosen. Tests on $WW$ simulation
show that the correct lepton was chosen in more than 99\% of the ambiguous
events.

Events not compatible with a 3-jet topology or with the lepton too close
to fragmentation products or inside a hadronic jet were recovered by looking
for particles inside jets with energy above 30~GeV and identified as electrons
or muons. In this case additional cuts were imposed on the impact
parameter of the lepton with respect to the beam spot and on the angles
which the missing momentum formed with respect to the beam direction and
the lepton itself.

The background contribution arising from the radiative return to the 
$Z$ peak was highly suppressed by rejecting events 
with the direction of the missing momentum 
close to the beam axis or with a detected photon
with an energy above 50~GeV. The cut on the polar angle of the missing momentum
was tighter for $q\bar{q}\tau\nu$ candidate events. 
The four-fermion neutral current background
($q\bar{q}l\bar{l}$) was suppressed by rejecting those events in which a 
second isolated and energetic lepton of the same flavour as the main 
candidate was found. Non-resonant contributions to the $q\bar{q}l\nu$
final state were reduced by requiring the invariant mass of the 
hadronic system to be larger than $20~{\mathrm GeV}/c^2$ 
and of the lepton-missing
momentum system to be larger than $10~{\mathrm GeV}/c^2$. 

The different leptonic decays were classified in the following way:
\begin{itemize}
\item $WW\rightarrow q\bar{q}\mu\nu$ 

The lepton was identified as a muon. The contamination from 
$q\bar{q}\tau\nu$ with $\tau \rightarrow \mu\nu\nu$ was suppressed by requiring
that, if the muon momentum was below $45~{\mathrm GeV}/c$, 
either the missing mass in the 
event was below $55~{\mathrm GeV}/c^2$,
or the fitted mass from a 2C kinematic 
fit with both $W$ bosons constrained to have the same mass was above
$75~{\mathrm GeV}/c^2$.
\item $WW\rightarrow q\bar{q}e\nu$ 

The lepton was identified as an electron. A cut on the aplanarity of the 
event, defined as the angle between the lepton direction and the plane 
of the two jets, was applied in order to reduce radiative and
non-radiative QCD background. The contribution from the process
$ee\rightarrow Zee$ was reduced by imposing requirements on the
invariant mass of the electron and the missing momentum (assumed to be 
a single electron in the beam pipe). A procedure identical to that
just described for the muon channel was then applied to reduce the 
contamination of $q\bar{q}\tau\nu$ events.
% in the selected electron sample.
\item $WW\rightarrow q\bar{q}\tau\nu$ 

The event was not classified as an electron or a muon decay. In order
to suppress the background from $e^+e^-$ annihilations into 
$q\bar{q}(\gamma)$, events containing a 1-prong candidate $\tau$ 
had to have aplanarity above $20\deg$.
For multi-prong $\tau$ events a cut on the effective centre-of-mass 
energy $\sqrt{s'}$ was added, requiring it to be between 105 and 
175~GeV. In both cases the hadronic system was rescaled to the beam energy
and cuts were applied to reject very low (less than $15~{\mathrm GeV}/c^2$)
and very high (more than $90~{\mathrm GeV}/c^2$) 
invariant masses in the resulting
jet-jet system or lepton-missing momentum system.
\end{itemize}

\begin{figure}[htb]
\centerline{\epsfig{file=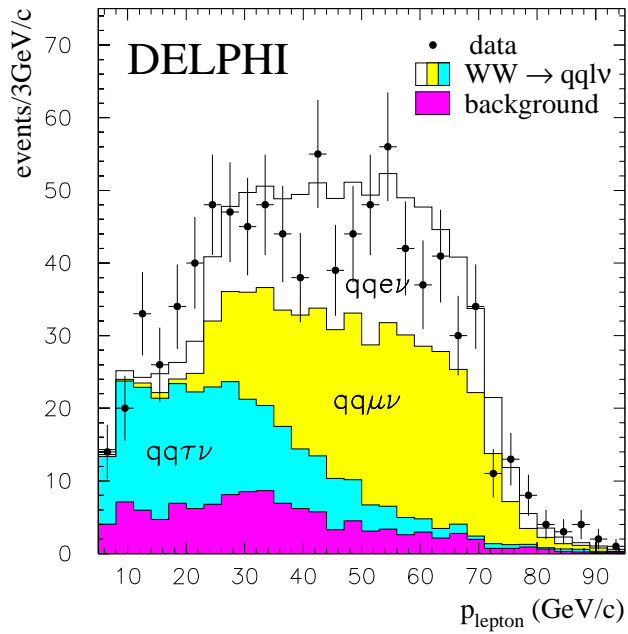,width=10cm}}
\caption{
Distribution of the lepton momentum for semileptonic events.
The points show the data and the histograms are the
predicted distributions for signal and background. The latter includes small
contributions from other $WW$ channels.
}
\label{fig2} \end{figure}

Figure~\ref{fig2} shows the distribution of the momentum of the selected 
leptons compared with the expectations from the simulation of signal and 
backgrounds. The numbers of selected events, efficiencies and backgrounds
for each lepton flavour are shown in Table~\ref{qqxx}. The efficiencies
include corrections to account for an imperfect description of the 
misidentification of electrons or muons as taus in the simulation,
for which an uncertainty of 1.0\% for electrons and 0.6\% for muons has been
taken into account in the determination of the $W$ branching ratios. 
The overall efficiency of the selection of $WW\rightarrow q\bar{q}l\nu$ events
was estimated to be $(75.4~\pm~0.7)$\%, varying significantly 
with lepton type: 92\% for $\mu$, 78\% for $e$ and
56\% for $\tau$ (see Table~\ref{qqxx}). 
The total expected background was estimated to be 
$(671~\pm~40)$~fb. The errors on efficiency and background include all
systematic uncertainties.

A total of 911 events were selected as semileptonic $W$ decays;
the number of events observed in the different lepton channels was found 
to be consistent with lepton universality. 
With the values given in Table~\ref{qqxx} for selected events, efficiencies
and backgrounds, and assuming lepton universality, a likelihood fit yields
a cross-section:
$$\sigma_{WW}^{l\nu qq} = \sigma_{WW}^{tot} 
\times BR(WW \rightarrow l\nu \qq) = 
6.77\pm0.26(\mbox{stat})\pm 0.12(\mbox{syst})~\mbox{pb}. $$
The systematic error includes contributions from efficiency and background,
four-fermion interference in the electron channel, imperfect 
track reconstruction description in the simulation 
and uncertainty in the luminosity measurement.

\subsection{Fully leptonic final state}

Events in which both $W$ bosons decay into $l\nu$ are characterized by low
multiplicity, a clean two-jet topology with two energetic, acollinear and
acoplanar leptons of opposite charge, and with large missing momentum and 
energy. The relevant backgrounds are di-leptons from 
$e^+e^- \rightarrow Z (\gamma)$, Bhabha scattering, two-photon collisions,
$Z e^+e^-$, $ZZ$ and single $W$ events.

The selection was performed in three steps. First a leptonic
preselection was made followed by $\mu$/$e$/$\tau$ identification
in both jets. Finally different cuts were applied for each channel 
to reject the remaining background, which was different in each case.

The leptonic preselection aimed to select a sample enriched in leptonic
events. All particles in the event were clustered into 
jets using the LUCLUS 
algorithm~\cite{LUCLUS} ($d_{join}=6.5~{\mathrm GeV}/c$) 
and only events with two
reconstructed jets, containing at least one charged particle each, were
retained. A charged particle multiplicity 
between 2 and 6 was required and at least one jet had to have
only one charged particle. 
The leading particle (that with the largest momentum) in each jet 
was required to have polar angle $\mid \cos\theta_l \mid < 0.98$. 
In order to reduce the background from two-photon collisions and 
radiative di-lepton events,
the event acoplanarity, $\theta_{acop}$, defined as the acollinearity of the 
two jet directions projected onto the plane perpendicular to the beam axis, 
had to be above $5\deg$. 
In addition, the total momentum transverse to the beam direction, $P_t$,
had to exceed 4\% of the centre-of-mass energy $\sqrt{s}$.
The associated electromagnetic energy for both leading particles was 
required to be less than $0.4\cdot \sqrt{s}$ to reject Bhabha scattering.

For this sample each particle was identified as $\mu$, $e$ or hadron.
Slightly different criteria for lepton identification were applied,
depending on whether the particle was in the barrel region 
($43^\circ<\theta<137^\circ$), in the forward region ($\theta<37^\circ$ or 
$\theta>143^\circ$), or in between. 
A particle was identified as a muon if at least one hit in the 
muon chambers was associated to it, or if it had deposited energy in the 
outermost layer of the hadron calorimeter; in addition the energy deposited
in the other layers had to be compatible with that from a minimum ionizing 
particle. 
For the identification of a particle as an electron the energies deposited 
in the electromagnetic calorimeters, in the different layers of the 
hadron calorimeter, and in addition the energy loss in the time projection 
chamber were used.
A lepton was identified as a cascade decay from $W \rightarrow \tau \nu_\tau$ 
if the momentum was lower than $20~{\mathrm GeV}/c$.

After the preselection and the channel identification, different 
cuts were applied depending on the channel in order to reject the remaining 
background. For all channels except $WW \rightarrow \mu \nu \mu \nu$, the
visible energy of the particles with $\mid \cos\theta \mid<0.9$ had to 
exceed $0.06 \cdot \sqrt{s}$. For all channels with at least one $W$ decaying
into $\tau \nu$, the invariant mass of each jet had to be below 
$3~{\mathrm GeV}/c^2$, the momentum of the leading particle of a candidate 
$\tau$ jet below $0.4 \cdot \sqrt{s}$, and $\theta_{acop}$ above $9\deg$.
In addition the following criteria were required for the individual channels:
\begin{itemize}
\item $WW\rightarrow e \nu e \nu$

The most important background comes from 
radiative Bhabha scattering. Therefore a cut on the 
neutral energy was imposed, and the acoplanarity had to be greater 
than $7^\circ$. In addition a minimum transverse energy was required and the
momenta of both leading particles had to be less than 45\% of the centre-of-mass
energy.

\item $WW\rightarrow \mu \nu \mu \nu$

In order to reject radiative di-muon events, the transverse energy 
was required to satisfy 
$0.2\cdot \sqrt{s} < E_{t} < 0.8 \cdot \sqrt{s}$, and the neutral energy had
to be less than 2~GeV.

\item $WW\rightarrow \tau \nu \tau \nu$

In order to reduce remaining background 
from $Z (\gamma)$-decays and from $\gamma\gamma \rightarrow \ell\ell$
processes, tighter cuts were applied on the acoplanarity, the
transverse momentum and the total transverse energy of the jets.
Finally the acollinearity was required to be between $10^\circ$ and
$150^\circ$.

\item $WW\rightarrow e \nu \mu \nu$

The neutral energy was required to be less than 20~GeV.

\item $WW\rightarrow \tau \nu e \nu$

A minimum transverse energy was required.
If one of the leading particles was not in the barrel region,
additional cuts on the momentum of each leading particle and on the acollinearity
were applied to reduce the background from two-photon collisions.

\item $WW\rightarrow \tau \nu \mu \nu$

The acoplanarity was required to be greater 
than $11^\circ$ if at least one leading particle was outside the barrel 
region.
\end{itemize}

\begin{figure}[htb]
\centerline{\epsfig{file=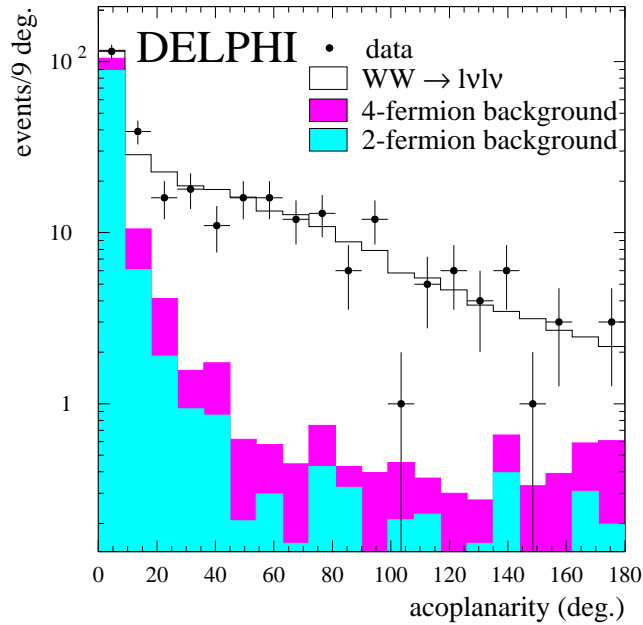,width=10cm}}
\caption{
Distribution of the acoplanarity angle for fully-leptonic events.
The points show the data and the histograms are the
predicted distributions for signal and background.}  
\label{fig3}
\end{figure}

The distribution of the acoplanarity angle, after having applied all cuts
except the one on the acoplanarity, is shown in Figure~\ref{fig3}.
The numbers of selected events, efficiencies and backgrounds in each channel are
shown in Table~\ref{lnulnu}. The overall $l\nu l\nu$ efficiency was 
($62.9 \pm 1.6$) \%. 
The efficiencies have been
corrected to account for an imperfect simulation of the misidentification of
electrons and muons as tau leptons in a similar way as for the semileptonic 
channel, and the uncertainty in the track reconstruction efficiency 
is taken into account in the total systematic error. Inefficiencies of the 
trigger are estimated to be $\leq 0.1\%$.
The residual background from non-$W$ and
single-$W$ events is $0.134 \pm 0.032$ pb, where the error includes 
all systematic effects introduced by the selection criteria.

\begin{table}[htb]
\begin{center}
\begin{tabular}{|c|c|c|c|c|c|c|c|}
\hline
& \multicolumn{6}{|c|}{efficiencies for selected channels} \\
\hline
channel              & $\tau \nu \tau \nu$ & $\tau \nu e \nu$ &
$\tau \nu \mu \nu$   & $e \nu e \nu$       & $e \nu \mu \nu$  & 
$\mu \nu \mu \nu$    \\ 
\hline
$\tau \nu \tau \nu$     & {\bf0.252} & 0.069 & 0.083 & 0.005 & 0.008 & 0.003 \\
$\tau \nu e \nu$        & 0.040 & {\bf0.433} & 0.012 & 0.044 & 0.057 & 0.    \\
$\tau \nu \mu \nu$      & 0.019 & 0.008 & {\bf0.540} & 0.0   & 0.043 & 0.047 \\
$ e \nu e \nu$          & 0.005 & 0.114 & 0.    & {\bf0.474} & 0.    & 0.    \\
$ e \nu \mu \nu$        & 0.004 & 0.038 & 0.090 & 0.001 & {\bf0.589} & 0.    \\
$ \mu \nu \mu \nu$      & 0.001 & 0.    & 0.058 & 0.    & 0.002 & {\bf0.655} \\
\hline
background (pb)         & 0.020 & 0.038 & 0.026 & 0.030 & 0.006 & 0.014 \\
selected events         & 15    & 40    & 43    & 20    & 38    & 27    \\
\hline
luminosity (pb$^{-1}$)  & \multicolumn{6}{|c|}{ 153.81 } \\ 
\hline
\end{tabular}
\vspace{0.2cm}
\caption{\label{lnulnu} Selection efficiencies, background and data for
the fully leptonic final states.}
\end{center}
\end{table}

A total of 183 events were selected in the data sample;
the number of events observed in the different di-lepton channels was found 
to be consistent with lepton universality. With the values given in
Table~\ref{lnulnu} for the selected events, efficiencies and backgrounds,
and assuming lepton universality, a likelihood fit yields a cross-section 

$$ \sigma_{WW}^{\ell\nu \ell\nu}= \sigma_{WW}^{tot} \times
   {\mathrm{BR}}(WW\rightarrow \ell\nu \ell\nu ) 
   = 1.68 \pm 0.14 ~\mbox{(stat)} \pm 0.07~\mbox{(syst)}
~~\mbox{pb}.  $$

The systematic error has contributions from the efficiency and background
determination, four-fermion interferences in the final states with electrons 
and from the measurement of the luminosity.

\section{Determination of total cross-section and branching fractions}

The total cross-section for $WW$ production 
and the $W$ leptonic branching fractions 
were obtained from likelihood fits to
%based on the probabilities of finding 
the numbers of events observed in each final state. 
The input numbers are those given in Tables~\ref{qqxx} 
and~\ref{lnulnu}, except for the fully hadronic final state where the binned
distribution of the neural network output was used.

From all the final states combined, the leptonic branching fractions with their
correlation matrix were obtained as shown in Table~\ref{brlept1}.
They are consistent with lepton universality. The fit was repeated
assuming lepton universality, and the results for the leptonic and derived
hadronic branching fraction are also given in Table~\ref{brlept1}.
The hadronic branching fraction is in agreement with the Standard Model 
prediction of 0.675. 

\begin{table}[p]
\begin{center}
\begin{tabular}{|l|c|c|c|c|}
\hline
channel&branching fraction&stat. error&syst. error & syst. from QCD bkg\\
\hline
$W \rightarrow e\nu$       &  0.1019 & 0.0064 & 0.0025 & 0.0005 \\
$W \rightarrow \mu\nu$     &  0.1076 & 0.0056 & 0.0012 & 0.0005 \\
$W \rightarrow \tau\nu$    &  0.1109 & 0.0087 & 0.0031 & 0.0003 \\
\hline
\end{tabular}
\vspace{0.3cm}
\begin{tabular}{|l|r r r|}
\hline
Correlations& $W \rightarrow e\nu$ & $W \rightarrow \mu\nu$ & $W \rightarrow \tau\nu$ \\
\hline
$W \rightarrow e\nu$    &  1.00 & -0.02 & -0.39 \\
$W \rightarrow \mu\nu$  & -0.02 &  1.00 & -0.29 \\
$W \rightarrow \tau\nu$ & -0.39 & -0.29 &  1.00 \\
\hline
\end{tabular}
\vspace{0.7cm}
\begin{tabular}{|l|c|c|c|c|}
\hline
& \multicolumn{4}{|c|}{assuming lepton universality} \\
\hline
channel&branching fraction& stat. error & syst. error & syst. from QCD bkg\\
\hline
$W \rightarrow \ell\nu$              & 0.1066 & 0.0028 & 0.0013 & 0.0004 \\
$W \rightarrow {\mathrm hadrons}$    & 0.6803 & 0.0084 & 0.0040 & 0.0013 \\
\hline
\end{tabular}
\vspace{0.2cm}
\caption{\label{brlept1} $W$ branching fractions from 189~GeV data
and correlation matrix for the leptonic branching fractions.
The uncertainty from the QCD background (column 5) is included in the 
systematic error (column 4).}
\end{center}
\end{table}

Assuming the other parameters of the Standard Model, i.e. elements $|V_{ud}|$, 
$|V_{us}|$, $|V_{ub}|$, $|V_{cd}|$ and $|V_{cb}|$ of the CKM matrix, lepton
couplings to $W$ bosons, and the
strong coupling constant $\alpha_S$, to be fixed at the
values given in~\cite{pdg}, the measured hadronic branching fraction 
can be converted \cite{Vcs} into
$$ |V_{cs}| = 1.001 \pm 0.040~\mbox{(stat)} \pm 0.020~\mbox{(syst)},$$
\noindent where the uncertainties of the Standard Model parameters are 
included in the systematic error.

The total cross-section for $WW$ production, with the assumption of 
Standard Model values for all branching fractions, was found to be 
$$ \sigma_{WW}^{tot} = 15.83 \pm 0.38~\mbox{(stat)} \pm 0.20~\mbox{(syst)}
~~\mbox{pb}.  $$

\begin{figure}[htb]
\centerline{\epsfig{file=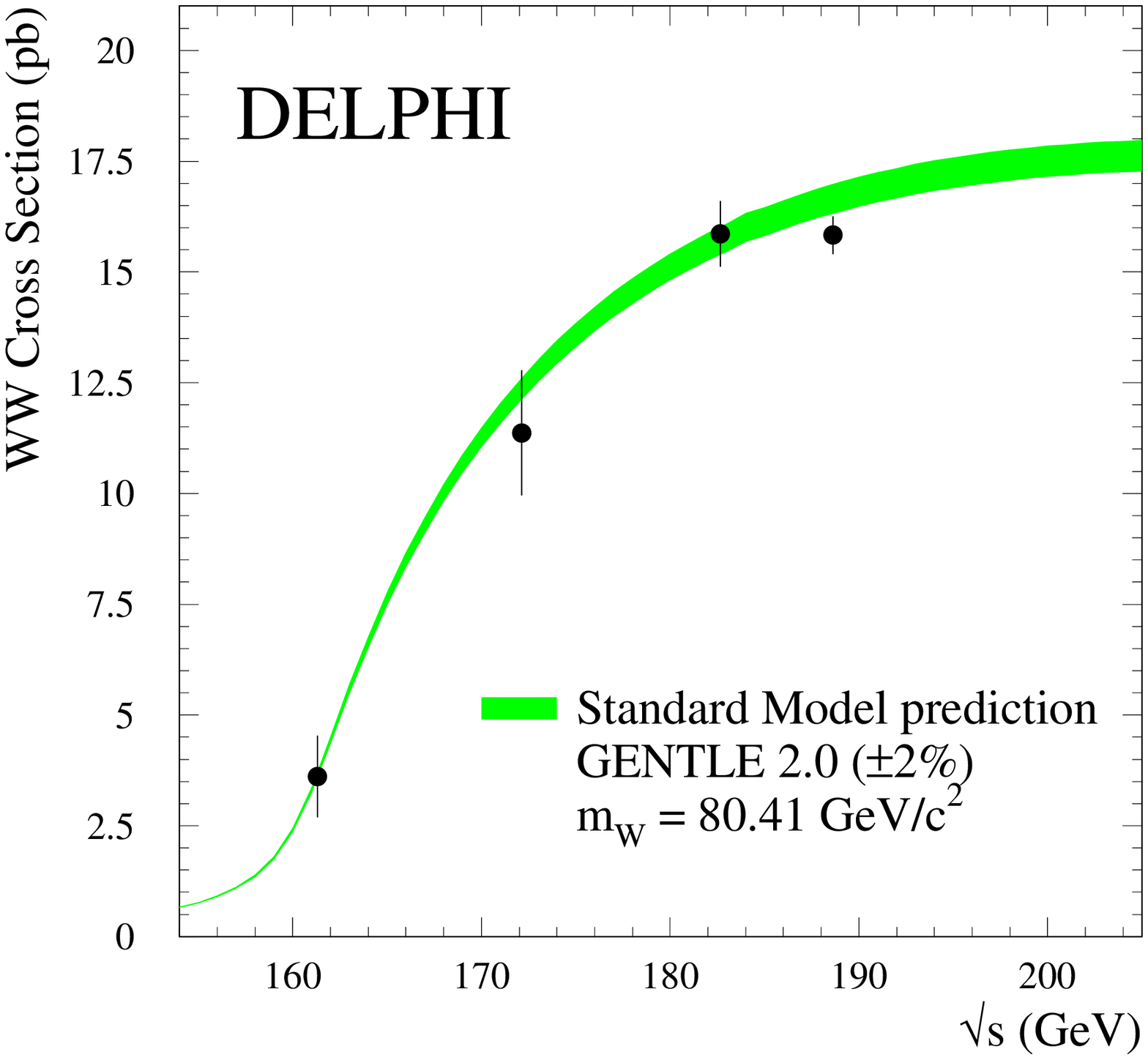,width=10cm}}
\caption{Measurements of the $W^+W^-$ cross-section 
compared with the Standard Model prediction using \protect\cite{gentle} and
$m_W = 80.41~{\mathrm GeV}/c^2$~\protect\cite{pdg}
with a possible uncertainty of $\pm 2\%$ on the computation.}
\label{fig4} \end{figure}

This result is shown in figure~\ref{fig4} together with the measurements
at lower centre-of-mass energies \cite{wwpap161,wwpap172,wwpap183}, 
and with the Standard Model prediction using GENTLE~\cite{gentle}.

The measurement of the branching fractions can be improved by combining the
present measurement with those at lower centre-of-mass 
energies~\cite{wwpap161,wwpap172,wwpap183}.
These results, obtained conservatively assuming full correlation of systematics
between different energies, are summarized in Table~\ref{brlept2}. 

\begin{table}[p]
\begin{center}
\begin{tabular}{|l|c|c|c|c|}
\hline
channel & branching fraction & stat. error & syst. error & syst. from QCD bkg\\
\hline
$W \rightarrow e\nu$       & 0.1018 & 0.0054 & 0.0026 & 0.0005 \\
$W \rightarrow \mu\nu$     & 0.1092 & 0.0048 & 0.0012 & 0.0006 \\
$W \rightarrow \tau\nu$    & 0.1105 & 0.0075 & 0.0032 & 0.0004 \\
\hline
\end{tabular}
\vspace{0.3cm}
\begin{tabular}{|l|r r r|}
\hline
Correlations& $W \rightarrow e\nu$ & $W \rightarrow \mu\nu$ & $W \rightarrow \tau\nu$ \\
\hline
$W \rightarrow e\nu$    &  1.00 & -0.02 & -0.38 \\
$W \rightarrow \mu\nu$  & -0.02 &  1.00 & -0.30 \\
$W \rightarrow \tau\nu$ & -0.38 & -0.30 &  1.00 \\
\hline
\end{tabular}
\vspace{0.7cm}
\begin{tabular}{|l|c|c|c|c|}
\hline
& \multicolumn{4}{|c|}{assuming lepton universality} \\
\hline
channel & branching fraction & stat. error & syst. error & syst. from QCD bkg\\
\hline
$W \rightarrow \ell\nu$                 &  0.1071 & 0.0024 & 0.0014 & 0.0005 \\
$W \rightarrow {\mathrm hadrons}$       &  0.6789 & 0.0073 & 0.0043 & 0.0015 \\
\hline
\end{tabular}
\vspace{0.2cm}
\caption{\label{brlept2} $W$ branching fractions from 
the combined 161, 172, 183 and 189~GeV data
and correlation matrix for the leptonic branching fractions.
The uncertainty from the QCD background (column 5) is included in the 
systematic error (column 4).}
\end{center}
\end{table}

\section{Summary}

From a data sample of $155 {\mathrm{~pb^{-1}}}$ integrated luminosity,
collected by DELPHI in \ee collisions
at a centre-of-mass energy of 188.63~GeV,
the individual leptonic branching fractions were found to be in agreement 
with lepton universality and
the $W$ hadronic branching fraction was measured to be
$${\mathrm{BR}}(W \rightarrow q\bar{q})
=0.680\pm 0.008({\mathrm stat}) \pm 0.004 ({\mathrm syst}),$$
in agreement with the Standard Model prediction of 0.675
and compatible with measurements at lower energies by other LEP 
experiments~\cite{aleph,l3,opal}.
The total cross-section for the doubly resonant $WW$ process was measured to be
$$\sigma_{WW}^{tot} = 15.83
\pm 0.38 ({\mathrm stat}) \pm 0.20 ({\mathrm syst})~{\mathrm pb},$$ 
assuming Standard Model branching fractions.

%=========================================================================%

\newpage

%         Modified on 04-06-1999 by dimartino
%-------------------------------------------------------------------
\subsection*{Acknowledgements}
\vskip 3 mm
We are greatly indebted to our technical 
collaborators, to the members of the CERN-SL Division for the excellent 
performance of the LEP collider, and to the funding agencies for their
support in building and operating the DELPHI detector.\\
We acknowledge in particular the support of \\
Austrian Federal Ministry of Science and Traffics, GZ 616.364/2-III/2a/98, \\
FNRS--FWO, Belgium,  \\
FINEP, CNPq, CAPES, FUJB and FAPERJ, Brazil, \\
Czech Ministry of Industry and Trade, GA CR 202/96/0450 and GA AVCR A1010521,\\
Danish Natural Research Council, \\
Commission of the European Communities (DG XII), \\
Direction des Sciences de la Mati$\grave{\mbox{\rm e}}$re, CEA, France, \\
Bundesministerium f$\ddot{\mbox{\rm u}}$r Bildung, Wissenschaft, Forschung 
und Technologie, Germany,\\
General Secretariat for Research and Technology, Greece, \\
National Science Foundation (NWO) and Foundation for Research on Matter (FOM),
The Netherlands, \\
Norwegian Research Council,  \\
State Committee for Scientific Research, Poland, 2P03B06015, 2P03B1116 and
SPUB/P03/178/98, \\
JNICT--Junta Nacional de Investiga\c{c}\~{a}o Cient\'{\i}fica 
e Tecnol$\acute{\mbox{\rm o}}$gica, Portugal, \\
Vedecka grantova agentura MS SR, Slovakia, Nr. 95/5195/134, \\
Ministry of Science and Technology of the Republic of Slovenia, \\
CICYT, Spain, AEN96--1661 and AEN96-1681,  \\
The Swedish Natural Science Research Council,      \\
Particle Physics and Astronomy Research Council, UK, \\
Department of Energy, USA, DE--FG02--94ER40817. \\
%=========================================================================%

%=========================================================================%
\end{document}